\def \SAIT #1 #2 {{\em Mem.\ Soc.\ Astron.\ It.\/} {\bf #1}, #2}
\def \MESS #1 #2 {{\em The Messenger\/} {\bf #1}, #2}
\def \ASTRNACH #1 #2 {{\em Astron. Nach.\/} {\bf #1}, #2}
\def \AAP #1 #2 {{\em Astron. Astrophys.\/} {\bf #1}, #2}
\def \AAL #1 #2 {{\em Astron. Astrophys. Lett.\/} {\bf #1}, L#2}
\def \AAR #1 #2 {{\em Astron. Astrophys. Rev.\/} {\bf #1}, #2}
\def \AAS #1 #2 {{\em Astron. Astrophys. Suppl. Ser.\/} {\bf #1}, #2}
\def \AJ #1 #2 {{\em Astron. J.\/} {\bf #1}, #2}
\def \ANNREV #1 #2 {{\em Ann. Rev. Astron. Astrophys.\/} {\bf #1}, #2}
\def \APJ #1 #2 {{\em Astrophys. J.\/} {\bf #1}, #2}
\def \APJL #1 #2 {{\em Astrophys. J. Lett.\/} {\bf #1}, L#2}
\def \APJS #1 #2 {{\em Astrophys. J. Suppl.\/} {\bf #1}, #2}
\def \APSS #1 #2 {{\em Astrophys. Space Sci.\/} {\bf #1}, #2}
\def \ASR #1 #2 {{\em Adv. Space Res.\/} {\bf #1}, #2}
\def \BAIC #1 #2 {{\em Bull. Astron. Inst. Czechosl.\/} {\bf #1}, #2}
\def \JSQRT #1 #2 {{\em J. Quant. Spectrosc. Radiat. Transfer\/} {\bf #1}, #2}
\def \MN #1 #2 {{\em Mon. Not. R. Astr. Soc.\/} {\bf #1}, #2}
\def \MEM #1 #2 {{\em Mem. R. Astr. Soc.\/} {\bf #1}, #2}
\def \PLR #1 #2 {{\em Phys. Lett. Rev.\/} {\bf #1}, #2}
\def \PASJ #1 #2 {{\em Publ. Astron. Soc. Japan\/} {\bf #1}, #2}
\def \PASP #1 #2 {{\em Publ. Astr. Soc. Pacific\/} {\bf #1}, #2}
\def \NAT #1 #2 {{\em Nature\/} {\bf #1}, #2}

\def\HST{{\sl HST\/ }}
\def\kms{\nobreak\mbox{$\;$km\,s$^{-1}$}}
\def\mag{\nobreak\mbox{$^{\rm m}$}\!\!\!\!.\;}

\documentstyle{memsait}
\input epsf.sty
\begin{opening}
\title{THE LUMINOSITY CALIBRATION OF SNe Ia: \\ PRESENT STATUS} 
\author{G.A. Tammamm \& B. Reindl}
\institute{Astronomisches Institut der Universit\"at Basel, \\
       Venusstr.~7, CH-4102 Binningen, Switzerland}
\date{} 
\end{opening}

\begin{document}

\ 
\bigskip

\begin{abstract}
Blue Supernovae of Type~Ia (SNe\,Ia) have become the most important
objects in cosmology being of exceptionally uniform luminosity.
Used as relative distance indicators they map deviations from pure
Hubble flow and determine the cosmological constant $\Lambda$. Once
their absolute magnitude is determined they provide the best estimate
of the large-scale value of the Hubble constant $H_0$. An \HST project
is reviewed where Cepheid distances are used for the luminosity
calibration of SNe\,Ia. The mean luminosity of 8 SNe\,Ia is 
$M_{\rm  B}(\max)=-19.47\pm0.07$, $M_{\rm  V}(\max)=-19.48\pm0.07$,
corresponding -- after small corrections for second parameters -- to
$H_0({\rm cosmic})=59\pm5$.
\end{abstract}

\section{Introduction}
As the number of SNe\,Ia increased and their data improved it became
increasingly clear that they are nearly perfect standard candles,
i.e. that their luminosity scatter at any given epoch (e.g. at
maximum) is $<0.2\;$mag
(Kowal 1968; 
 Sandage \& Tammann 1982,~1993; 
 Tammann 1979, 1982;
 Cadonau, Sandage, \& Tammann 1985;
 Branch \& Tammann 1992; 
 Tammann \& Sandage 1995; 
 Hamuy et~al. 1995, 1996a),
if one excludes the few red SNe\,Ia, which are underluminous because
of absorption and/or intrinsic peculiarities, and restricts oneself to
blue, spectroscopically uniform (``Branch-normal'';
Branch, Fisher, \& Nugent 1993) SNe\,Ia with $(B_{\max} -V_{\max})\le 0.2$.
The scatter becomes even smaller when small differences in decline
rate and color are allowed for (Parodi et~al. 1999, and sources
therein). This gives SNe\,Ia a unique role in cosmology. Their
apparent magnitudes can be used to map deviations from pure Hubble
flow and hence to trace large-scale inhomogeneities
(Tammann 1998a, 1999; Zehavi et~al. 1998), as well as to determine
the value of the cosmological constant $\Lambda$
(Perlmutter et~al. 1999; Schmidt et~al. 1998); in the latter case one
depends, of course, on the absence of secular luminosity evolution
effects (cf. Kobayashi et~al. 1998).

In addition, if it is possible to calibrate the luminosity of SNe\,Ia
in absolute terms, one can derive unparalleled distances to better
than 10\% to all galaxies which have produced well observed SNe\,Ia
and determine the Hubble constant $H_0$ out to velocity distances
of $30\,000\kms$. This distance range is ideally suited because one is
bound to find here the truly cosmic value of $H_0$, unperturbed
by peculiar velocities, and yet the K-correction is still small and
space curvature effects negligible.

   For the luminosity calibration of SNe\,Ia a small \HST team
has been formed comprising A.~Sandage, A.~Saha, L.~Labhardt,
F.\,D.~Macchetto, N.~Panagia, \& G.\,A.~Tammann. 
They have observed so far the Cepheids in six galaxies having produced
seven SNe\,Ia with good or even excellent light curve coverage. 
The resulting Cepheid distances yield the absolute magnitudes 
$M_{\rm B}$ and $M_{\rm V}$ of the SNe\,Ia at maximum light.
Their results shall be reported in the following. The reader is also
referred to other reviews on the subject (Saha 1998, 1999;
Macchetto \& Panagia 1999).

The reliability of Cepheids is briefly discussed in Sec.~2. The
results concerning SNe\,Ia are compiled in Sec.~3. A discussion of
second-parameter effects is found in Sec.~4. The conclusions are
in Sec.~5.

\section{Cepheids as Distance Indicators}
Cepheids are presently, through their period-luminosity (PL)
relation, the most reliable and least controversial distance
indicators. The slope and the zeropoint of the PL relation is taken
from the very well-observed Cepheids in the Large Magellanic Cloud
(LMC), whose distance modulus is adopted to be $(m-M) = 18.50$
(Madore \& Freedman 1991).

   An old PL relation calibrated by Galactic Cepheids in open cluster,
and now vindicated by Hipparcos data (Sandage \& Tammann 1998), gave
$(m-M)_{\rm LMC}=18.59$ (Sandage \& Tammann 1968, 1971). Hipparcos data
combined with more modern Cepheid data give an even somewhat higher modulus
(Feast \& Catchpole 1997). Reviews of Cepheid distances (Federspiel,
Tammann, \& Sandage 1998; Gratton 1998) cluster around $18.56\pm0.05$,
-- a value in perfect agreement with the purely geometrical distance
determination of SN\,1987A ($18.58\pm0.05$; Gilmozzi \& Panagia
1999). In his excellent review Gratton (1998) concludes from the rich
literature on RR\,Lyr stars that $(m-M)_{\rm LMC}=18.54\pm0.12$. He
also discusses five distance determination methods which give lower
moduli by $0.1-0.2\;$mag, but they are still at a more experimental
stage. -- There is therefore emerging evidence that the adopted LMC
modulus of 18.50 is too small by $\sim\!0.06\;$mag and that the
SNe\,Ia luminosities derived in the next paragraph should be increased
by this amount.

   There has been much debate about the possibility that the PL
relation of Cepheids depends on metallicity. The question is here
of more principal than practical importance because the {\it mean\/}
metallicity of the seven spiral galaxies and one Am galaxy considered
below hardly differs by much from the Galactic metallicity. Direct
{\it observational\/} evidence for a (very) weak metallicity
dependence comes from the fact that the metal-rich
Galactic Cepheids give perfectly reasonable distances for the
moderately metal-poor LMC Cepheids and the really metal-poor SMC
Cepheids and, still more importantly, that their relative distances
are wavelength-independent (Di Benedetto 1997; cf. Tammann 1997). 
-- Much progress has been made on the theoretical front. Saio
\& Gautschy (1998) and Baraffe et~al. (1998) have evolved Cepheids
through the different crossings of the instability strip and have
investigated the pulsational behavior at any point. The resulting
(highly metal-insensitive) PL relations in bolometric light have been
transformed into PL relations at different wavelengths by means of
detailed atmospheric models; the conclusion is that any metallicity
dependence of the PL relations is negligible (Sandage, Bell, \&
Tripicco 1998; Alibert et~al. 1999; cf. however Bono, Marconi, \&
Stellingwerf 1998, who strongly depend on the treatment of stellar
convection). 

   While remaining uncertainties of the PL relation and the zeropoint
seem to have only minor practical consequences, the application to
\HST observations is by no means simple. The photometric zeropoint,
the linearity over the field, crowding, and cosmic rays raise
technical problems. The quality of the derived distances depends
further on (variable) internal absorption and the number of available
Cepheids in view of the finite width of the instability strip. (An
attempt to beat the latter problem by using a PL-color relation is
invalid because the underlying assumption of constant slope of the
constant-period lines is unrealistic; cf. Saio \& Gautschy
1998). Typical errors of individual Cepheid distances from \HST are
therefore $\pm0.2\,$mag (10\% in distance). For five of the nine
SNe\,Ia in Table~1 (below) the resulting errors in luminosity are
smaller, because they suffer closely the same (small) absorption as
``their'' Cepheids such that only apparent distance moduli are needed.

\section{The Calibration of SNe\,Ia Luminosities from Cepheid
  Distances}
The \HST project for the luminosity calibration of SNe\,Ia has so far
provided Cepheid distances for six galaxies which have produced seven
SNe\,Ia. Their absolute $B$ and $V$ magnitudes at maximum are repeated
here in Table~1 from Parodi et~al. (1999). The individual errors are
compounded from the error of the Cepheid distance and the goodness of
the SN light curve; they are listed as well as the original archive
sources by Saha et~al. (1999). On average the errors amount to
$0\mag22$ in B and $0\mag20$ in $V$.




%
\begin{table}[t]
\begin{center}
\small\rm
\caption{Absolute $B$ and $V$ magnitudes of blue SNe\,Ia calibrated
  through Cepheid distances of their parent galaxies}
\label{tab:1}
\begin{minipage}{0.91\textwidth}
\begin{tabular}{lllrrrrc}
\noalign{\smallskip}
\hline
\noalign{\smallskip}
  SN & Galaxy & $\log v_{220}^{1)}$ & 
  \multicolumn{1}{c}{m$_{\rm B}^{5)}$} & 
  ($B-V$)$^{7)}$& M$_{\rm B}$ &  M$_{\rm V}$ & $\Delta m_{15}$ \\
\noalign{\smallskip}
\hline
\noalign{\smallskip}
1895B  & NGC\,5253  & 2.464$^{2)}$ &  8.26\hspace*{7.5pt} &  
 \multicolumn{1}{c}{$\cdots$} & -19.87 & 
 \multicolumn{1}{c}{$\cdots$} & 
 \multicolumn{1}{c}{$\cdots$} \\
1937C  & IC\,4182   & 2.519        &  8.83\hspace*{7.5pt} & -0.05\hspace*{11pt} &
-19.53 & -19.48 & 0.87 \\
1960F  & NGC\,4496A & 3.072$^{3)}$ & 11.60\hspace*{7.5pt} &  0.06\hspace*{11pt} &
-19.56 & -19.62 & \multicolumn{1}{c}{$\cdots$} \\
1972E  & NGC\,5253  & 2.464$^{2)}$ &  8.61\hspace*{7.5pt} & -0.03\hspace*{11pt} &
-19.52 & -19.49 & 0.87 \\
1974G  & NGC\,4414  & 2.820        & 11.82$^{6)}$ &  0.02\hspace*{11pt} & -19.59 &
-19.61 & 1.11 \\
1981B  & NGC\,4536  & 3.072$^{3)}$ & 11.64$^{6)}$ & -0.02\hspace*{11pt} & -19.46 &
-19.44 & 1.10 \\
1989B  & NGC\,3627  & 2.734        & 10.86$^{6)}$ & -0.02\hspace*{11pt} & -19.36 &
-19.34 & 1.31 \\
1990N  & NGC\,4639  & 3.072$^{3)}$ & 12.76\hspace*{7.5pt} &  0.06\hspace*{11pt} &
-19.27 & -19.33 & 1.03 \\
1998bu & NGC\,3368  & 2.814$^{4)}$ & 10.84$^{6)}$ & -0.02\hspace*{11pt} & -19.53 &
-19.51 & 1.01 \\
\noalign{\smallskip}
\hline
\noalign{\smallskip}
 \multicolumn{4}{l}{mean (straight), excluding SN\,1895B} 
                                    &     & -19.48 & -19.48 & \\ 
 & & & & & $\pm .04$ & $\pm 0.04$ & \\
 \multicolumn{4}{l}{mean (weighted), excluding SN\,1895B} 
                                    &     & -19.47 & -19.48 & \\
 & & & & & $\pm .07$ & $\pm 0.07$ & \\
\noalign{\smallskip}
\hline
\noalign{\smallskip}
\end{tabular}

{\footnotesize
 $^{1)}$The velocities used are corrected for Virgocentric infall
 assuming a local infall velocity of $220\kms$.\par
 $^{2)}$The mean velocity $v_{220}=291\kms$ of the CenA group is
 used.\par
 $^{3)}$The mean velocity $v_{220}=1179\kms$ of the Virgo cluster is
 used.\par 
 $^{4)}$The mean velocity $v_{220}=652\kms$ of the Leo group is
 used.\par
 $^{5)}$The sources for the apparent magnitudes, corrected for
 Galactic absorption, are given by Parodi et~al. 1999.\par
 $^{6)}$Corrected for additional absorption in the parent galaxy.\par
 $^{7)}$Corrected for galactic and internal absorption where applicable.
}
\end{minipage}
\end{center}
\end{table}

Table~1 contains two additional SNe\,Ia. A Cepheid distance of
NGC\,4414 has been published by Turner et~al. (1998). The photometric
data of the corresponding SN\,Ia, SN\,1974G, have been compiled by
Schaefer (1998). SN\,1998bu has occurred in NGC\,3368, for which a
Cepheid distance was already available (Tanvir et~al. 1995); the
photometry of the SN is excellent (Suntzeff et~al. 1998).

   The quality of the four reddest SNe\,Ia in Table~1 is somewhat
downgraded by corrections for internal absorption which is clearly
present and larger than for the Cepheids in the same galaxy. This
affects particularly SN\,1974G, SN\,1989B, and SN\,1998bu. The
different sources of the color excesses $E(B-V)$ are given by Saha
et~al. (1999).  

   A glance at Table~1 shows that the old SN\,1895B was somewhat
overluminous. It is omitted in the following mainly because its
$V$-magnitude is not known. The scatter in absolute magnitude of the
remaining eight SNe\,Ia amounts to only $0\mag11$, which is
considerably less than the {\em estimated\/} individual errors and
provides independent proof of SNe\,Ia being powerful standard candles.

   The mean absolute magnitudes at B and V maximum of the eight
SNe\,Ia are shown in Table~1, weighted and unweighted. We adopt in
the following
\begin{eqnarray}
  \label{eq:1}
  M_{\rm B}(\max) & = & -19.47 \pm 0.07, \quad {\rm and} \\
  \label{eq:2}
  M_{\rm V}(\max) & = & -19.48 \pm 0.07.
\end{eqnarray}

   The empirical luminosity calibration is in perfect agreement with
presently available theoretical models. H\"oflich \& Khokhlov (1996)
have fitted sufficiently blue models, i.e. $(B-V)<0.2$ at maximum, to
the light curves and spectra of 16 SNe\,Ia. Their mean luminosity is
identical to equations~(1) and (2). In a recent review Branch (1998)
has concluded that present theory is best satisfied by $M_{\rm B}
\approx M_{\rm V}=-19.4$ to $-19.5$ for blue SNe\,Ia.

   Della Valle et~al. (1998) have attempted to derive an independent
distance of NGC 1380, host of SN\,1992A, by means of the peak of
the luminosity function of globular clusters (GCLF) and advocated a
low luminosity of the SN. However, the GCLF method is known to give
sometimes erratic results (Tammann 1998b).

   Kennicutt, Mould, \& Freedman (1998), and Freedman (1999) have
discarded several of the calibrators in Table~1 and added two that are
not based on direct Cepheid distances to the host galaxy. They have
consequently derived a fainter mean absolute $B$ magnitude than
equations~(1) and (2). Specifically, they assume that the distance of
the early-type galaxies NGC\,1316 and NGC\,1380 in the Fornax cluster,
parent galaxies of SN\,1980N and SN\,1992A, are identical with that of
the spiral NGC\,1365 for which there is a Cepheid distance. Suntzeff
et~al. (1998) have also considered the questionable SN\,1980N and
SN\,1992A as possible calibrators.
 
   However, there are reasons to suspect that NGC\,1365 is in the
foreground of the Fornax cluster and therefore that the precept of the
fainter calibration used by Kennicutt et~al. (1998),
calibrating the two Fornax SNe\,Ia via NGC\,1365, is not correct. The
evidence is that Wells et~al. (1994) have demonstrated that the
multi-color light curves of SN\,1989B in NGC\,3627 and SN\,1980N in
NGC\,1316 are virtually identical, and in fact establish the reddening
and extinction to SN\,1989B by comparing the magnitude shifts in
different passbands relative to SN\,1980N. Asserting then that
SN\,1980N has the same peak brightness as SN\,1989B yields the
distance modulus difference of $1\mag62\pm0\mag03$ (Wells
et~al. 1994). There is additional uncertainty of $\pm0.17$ to allow
for scatter in the difference in peak brightness of two SNe\,Ia with
the same decline rate (cf. equation~(5) below). With the Cepheid modulus of
$(m-M)_0=30.22\pm0.12$ (Saha et~al. 1999) for SN\,1989B, the derived
modulus of NGC\,1316, host to SN\,1980N, is $31.81\pm0.21$ (cf. also
Fig.~2 below). This is $0\mag5$ more distant than the Cepheid distance
of NGC\,1365 (Madore et~al. 1998), but is closely the same as found
for the early-type galaxies of the Fornax cluster by independent
methods (Tammann 1998b). In any case the implication of Kennicutt
et~al. (1998) and Freedman (1999) that the two {\em twin\/}
SNe\,Ia 1989B and 1980N should differ by $0\mag5$ in luminosity has little
credibility. 

   Some advocates of a high value of $H_0$ have excluded the brighter
calibrators in Table~1 for various arguments. If they are replaced by
SN\,1980N and SN\,1992A, for which no direct Cepheid distances exist,
the mean luminosity of blue SNe\,Ia becomes arbitrarily low (and the
value of $H_0$ equally high). 

   The \HST project for the luminosity calibration of SNe\,Ia will be
continued. Cepheid observations are presently granted for NGC\,4527
(with the peculiar-spectrum and possibly overluminous SN\,1991T)
and NGC\,3982 (with SN\,1998aq).

\section{SNe\,Ia Luminosities in Function of Second Parameters}
There is a rich literature on distance-independent observables which
govern the luminosity of SNe\,Ia. The original driver was the hope to
unify the bright, blue SNe\,Ia with ``Branch-normal'' spectra  and the
faint, red SNe\,Ia with peculiar spectra. 
The latter are easily distinguished by the observer and are
here excluded a priori, because blue and red SNe\,Ia may well form a
dichotomy (e.g. Nadyozhin 1998). But even for the blue SNe\,Ia with a
total luminosity range of $\sim\!0\mag6$ it is justified to ask
whether their luminosity depends on second parameters.

   Indeed it could be established that the luminosity of blue SNe\,Ia
correlates with the decline rate $\Delta m_{15}$ (or some other
measure of the light curve shape), the SN color, and the Hubble type
(or color) of the parent galaxy. There must be other more physical
parameters like Ni mass, temperature, expansion velocity, or spectral
features which correlate with luminosity, but the data are presently
too sparse to be useful. An increase of the luminosity scatter with
increasing galactocentric distance, proposed by Wang, H\"oflich, \&
Wheeler (1997), does not seem to be significant for blue SNe\,Ia
(Parodi et~al. 1999).

   The difficulty of establishing a SN luminosity - second-parameter
correlation is that independently determined luminosities are
needed. All known distance indicators with the necessary range, except
Cepheids observed with {\sl HST}, are much too inaccurate for the
purpose. Since only {\em relative\/} luminosities and hence distances
are needed, regress was taken to velocity distances. But even beyond
$v=1000\kms$ peculiar motions and a possible ``overexpansion'' out to
$10\,000\kms$ (Tammann 1998a, 1999; Zehavi et~al. 1998) may affect the
luminosities derived from recession velocities. It is safe therefore
to concentrate on blue SNe\,Ia beyond $10\,000\kms$.

   The best visualization of how to derive second-parameter
correlations is provided by the Hubble diagram of blue SNe\,Ia
(Fig.~1). Here the 54 blue SNe\,Ia with known $B$ and $V$ maxima, are
plotted, as compiled by Parodi et~al. (1999) from the Tololo/Calan
survey (Hamuy et~al. 1996b) and other sources. Correlations of the
residuals (read in magnitude) from the mean Hubble line with slope 0.2
can now be seeked with any second parameters, giving highest weight to
SNe\,Ia with $v>10\,000\kms$ because of non-linearity effects of the
expansion field. A detailed analysis is given by Parodi
et~al. (1999). Here only a brief summary is given.

\begin{figure}
\epsfxsize=8cm 
\hspace{2.4cm}\epsfbox{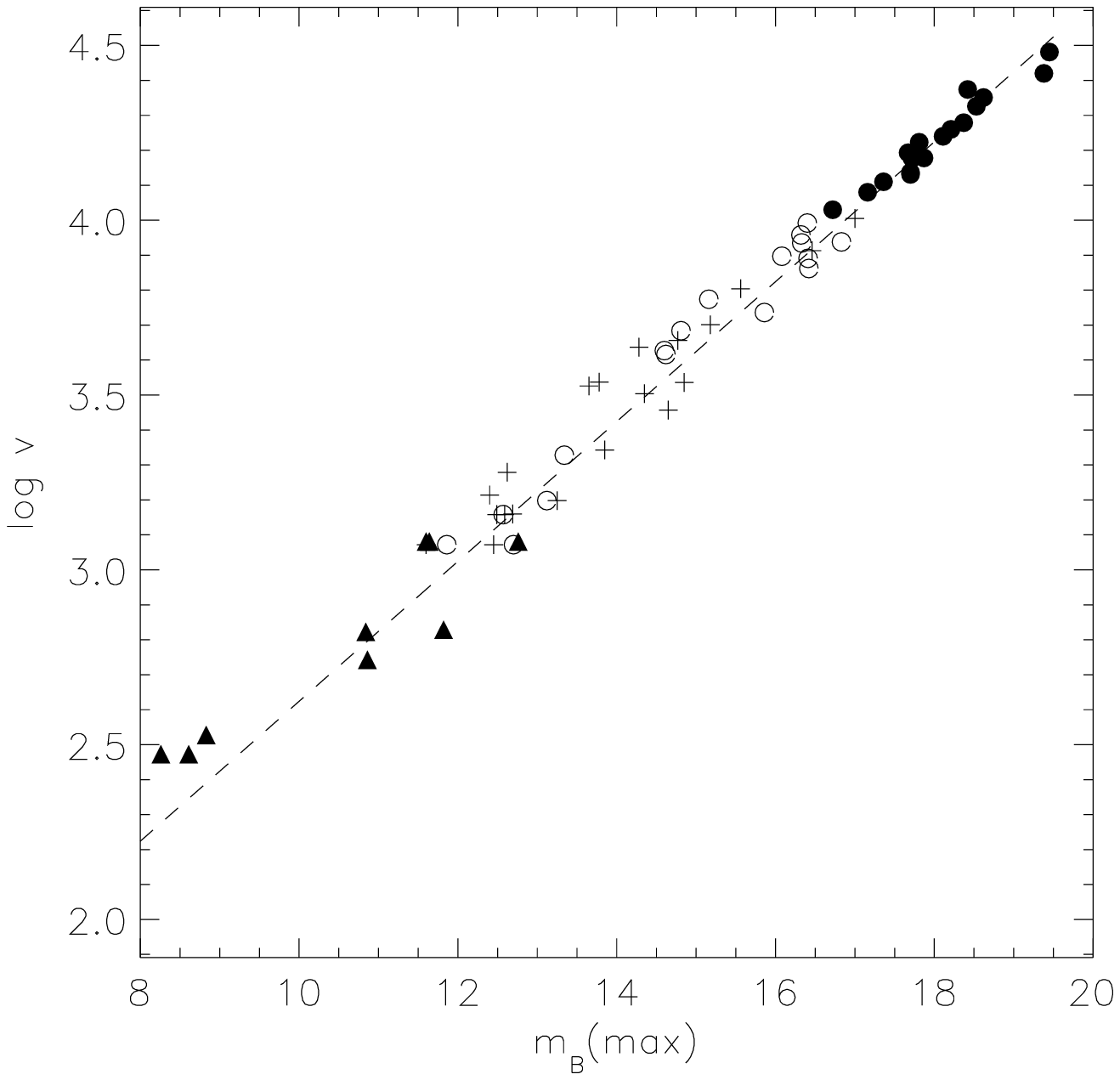} 

\epsfxsize=8cm 
\hspace{2.4cm}\epsfbox{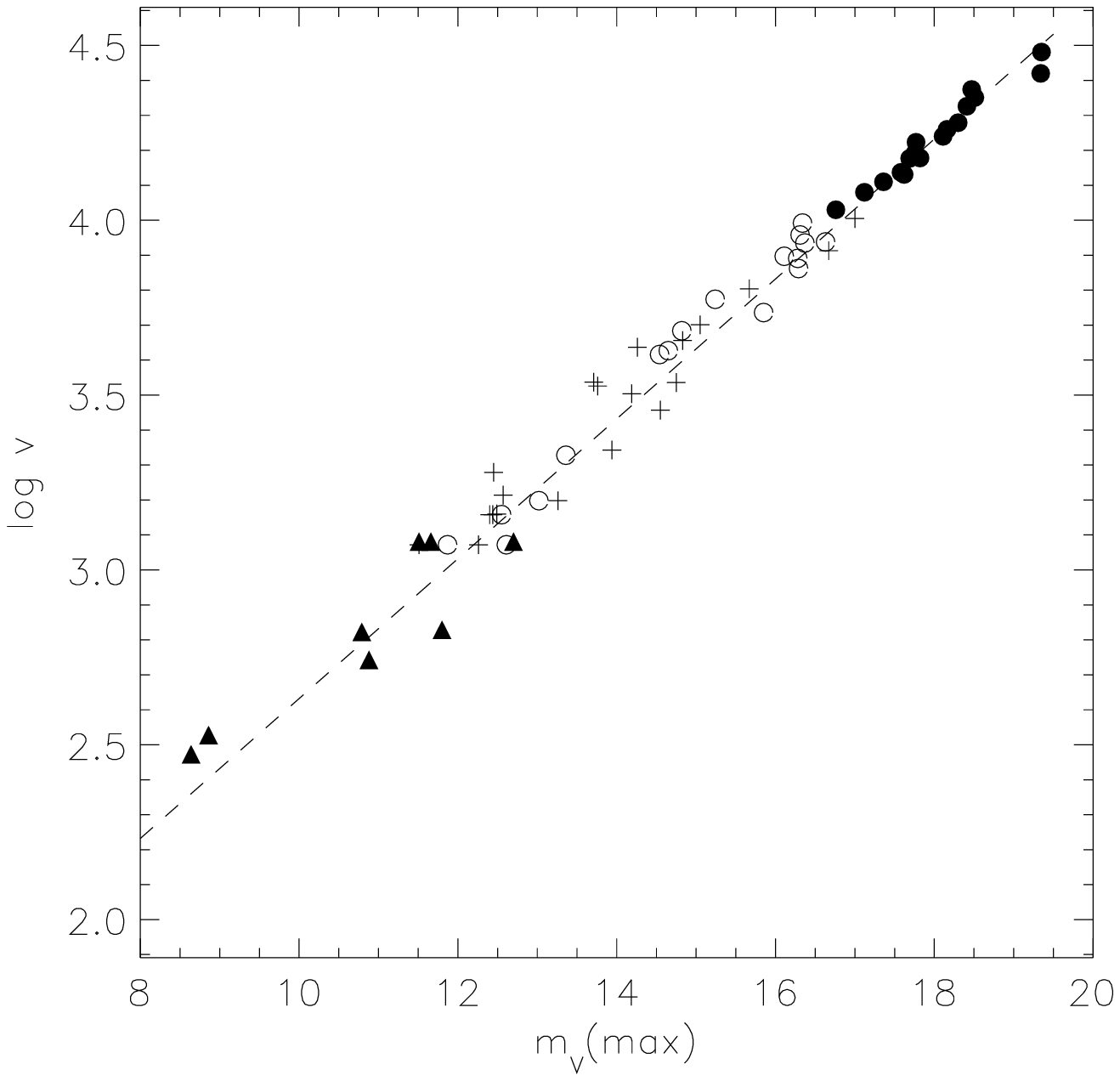}
\caption[h]{(a) The Hubble diagram in $B$ for 54 blue SNe\,Ia. The
  data before 1985 are represented with crosses. High-quality SNe\,Ia
  after 1985 with $v<10\,000\kms$ are shown with open circles, the more distant
  ones with filled circles. The eight calibrating SNe\,Ia are shown
  with triangles for comparison. (b) The same in $V$ magnitudes. The
  fitted lines have slope 0.2 according to equations~(3) and (4); they
  only consider SNe\,Ia with $v>10\,000\kms$. -- The velocities are
  corrected for a self-consistent Virgocentric infall model
  (Kraan-Korteweg 1986) if $v<3000\kms$; for $v>3000\kms$ an
  additional correction was applied for the motion of $630\kms$
  relative to the CMB dipole anisotropy (Smoot et~al. 1992). -- (From
  Parodi et~al. 1999).}
\end{figure}

   The 17 SNe\,Ia with $v>10\,000\kms$ define the Hubble line with
forced slope 0.2 as follows 
\begin{eqnarray}
  \label{eq:3}
  \log v & = & 0.2 m_{\rm B} + (0.62\pm0.01), \quad \sigma_{\rm B} = 0.18 \\
  \label{eq:4}
  \log v & = & 0.2 m_{\rm V} + (0.63\pm0.01), \quad \sigma_{\rm V} = 0.15.
\end{eqnarray}
The small scatter, before any second-parameter correction is applied,
should be noted.

   The best correlation between the luminosity and the decline rate
$\Delta m_{15}$ is given by the same SNe\,Ia as (assuming arbitrarily
$H_0=55$)
\begin{eqnarray}
  \label{eq:5}
  M_{\rm B} & = & -19.56 + 0.62 (\Delta m_{15} - 1.1), \quad \sigma_{\rm B}
  = 0.17 \\
  \label{eq:6}
  M_{\rm V} & = & -19.59 + 0.49 (\Delta m_{15} - 1.1), \quad \sigma_{\rm V} = 0.13,
\end{eqnarray}
where the scatter is only marginally reduced.

   The luminosity-SN\,color relation, again for the same SNe\,Ia, is
given by
\begin{equation}
  \label{eq:7}
  M_{\rm B} = -19.58 + 1.81 (B_{\max} - V_{\max}), \quad \sigma_{\rm B} = 0.15.
\end{equation}
The corresponding equation for $M_{\rm V}$ follows trivially by
decreasing the slope by one unit. SN~color is apparently somewhat more
efficient in reducing the scatter than the decline rate $\Delta
m_{15}$.

   In principle the variation in color can be intrinsic to the present
sample of SNe\,Ia or can be caused by absorption in the host
galaxy. Riess, Press, \& Kirshner (1996), Riess et~al. (1998), and
Phillips et~al. (1999) have argued that absorption is the main
cause. However, in that case the slope in equation~(7) would have to
be close to 4. A definite proof that {\em most\/} of the color
variation is in fact {\em intrinsic}, which is also expected from all
SN models, has been provided by Saha et~al. (1999) who have shown that
SNe\,Ia become {\em brighter\/} in $I$ as $(B_{\max}-V_{\max})$
becomes {\em redder}, and again {\em brighter\/} in $V$ and $I$ as
$(V_{\max}-I_{\max})$ becomes {\em redder}!

  The fact that internal absorption plays a relatively minor role in
distant SNe\,Ia, whereas four of the calibrating SNe\,Ia suffer an
appreciable absorption is not as puzzling as it may appear at first
sight, because observational selection effects dominate the available
samples. The apparent magnitudes of the nearby calibrators are so
bright that some absorption hardly hampers their discovery. At large
distances, however, only the {\em apparently\/} very brightest SNe\,Ia
are discovered, which automatically discriminates against absorption.

   Present data do not allow to derive a clear-cut correlation between
luminosity and Hubble type T. While the nearer SNe\,Ia in
ellipticals are $\sim\!0\mag2$ fainter in E/S0 galaxies than in
spirals (note that any absorption correction could only increase the
difference), the difference disappears beyond $10\,000\kms$. Clearly
larger samples are needed. In any case the Hubble type T is strongly
correlated with $\Delta m_{15}$, and once the SN luminosities are
corrected for $\Delta m_{15}$ and color no significant correlation
with T remains.

   If the 17 most distant SNe\,Ia of the sample are reduced to $\Delta
m_{15}=1.1$ and $(B_{\max}-V_{\max})=0.00$ by means of
equations~(5)--(7) one obtains a corrected Hubble line of form
\begin{equation}
  \label{eq:8}
  \log v = 0.2 m_{\rm B}^{\rm corr} + (0.656\pm0.005), \quad \sigma_{\rm
  B} = 0.12. 
\end{equation}
Here the small scatter is close to what can be expected from the
observational errors in $m_{\rm B}$, $m_{\rm V}$, and $\Delta
m_{15}$. Any additional reduction of the scatter would therefore be
artificial. (The two-step procedure to correct first for $\Delta
m_{15}$ and then for color is statistically not correct. However,
$\Delta m_{15}$ and color are so weakly correlated that a simultaneous
solution gives exactly the same result [cf. Parodi et~al. 1999]).

\section{Conclusions}
The most interesting conclusion for the physicist is that the
empirical luminosity calibration of SNe\,Ia is in very good agreement
with the results of present model calculations.

   The cosmologist must be pleased that uncorrected blue SNe\,Ia have
a luminosity scatter of only $\sigma_{\rm B} = 0.18$ and $\sigma_{\rm
  V} = 0.15$ (and even less in $I$). After correction for $\Delta
m_{15}$ (or any similar light curve shape parameter) and color the
scatter is reduced to $\sigma_{\rm B} = 0.12$, which might entirely be
caused by observational errors. It is not likely that the Universe
will ever offer better standard candles.

   For the astronomer blue SNe\,Ia provide an unparalleled tool for
the derivation of individual galaxy distances and of the large-scale
value of $H_0$.

   Equations~(3) and (4) can easily be transformed into
\begin{eqnarray}
  \label{eq:9}
  \log H_0 (B) & = &  0.2 M_{\rm B} + (5.62 \pm 0.01), \quad {\rm and} \\   
  \label{eq:10}
  \log H_0 (V) & = &  0.2 M_{\rm V} + (5.63 \pm 0.01).
\end{eqnarray}
Inserting the absolute magnitudes from equations~(1) and (2)
immediately yields
\begin{eqnarray}
  \label{eq:11}
  H_0 (B) & = & 53.2 \pm 2.0 \quad ({\rm internal\; error}), \\
  \label{eq:12}
  H_0 (V) & = & 54.2 \pm 2.0 \quad ({\rm internal\; error}).
\end{eqnarray}
This value holds for SNe\,Ia beyond $10\,000\kms$.

   Correspondingly the SNe\,Ia corrected for $\Delta m_{15}$ and color
give from equation~(8)
\begin{equation}
  \label{eq:13}
   \log H_0 (B,V) =  0.2 M_{\rm B}^{\rm corr} + (5.656 \pm 0.015).
\end{equation}
The $B$ and $V$ magnitudes lead to the same relation by construction
because they are reduced to $(B_{\max}-V{\max})=0.00$. -- If the
analogue corrections (from equations~(5)--(7)) are applied to 
the seven calibrators in Table~1 with known $\Delta m_{15}$, one
obtains
\begin{equation}
  \label{eq:14}
  M_{\rm B}^{\rm corr} = -19.44 \pm 0.04,
\end{equation}
and from a combination of equations~(13) and (14)
\begin{equation}
  \label{eq:15}
  H_0 ({\rm cosmic}) = 58.6 \pm 1.2 \quad({\rm internal\; error}).
\end{equation}

   The value of $H_0$ after correction for $\Delta m_{15}$ and color
is 9\% larger than without these corrections. The reason is that the
local calibrators are somewhat slower decliners ($\delta\Delta
m_{15}=0.29$) and somewhat bluer ($\delta(B-V)=0.04$) than the
distant SNe\,Ia. This in turn is caused by the requirement that the
calibrators have to be in spirals (with Cepheids) while the distant
objects come in galaxies of all kinds of Hubble types. 

\begin{figure}
\begin{center}
{\footnotesize \sf
\def\floatwidth{0.25\textwidth}
\setlength{\unitlength}{1mm}
\def\xx{120}
\def\yy{151}
\begin{picture}(\xx,\yy)(0,0)
 \put(0,0){\line(\xx,0){\xx}}
 \put(0,0){\line(0,\yy){\yy}}
 \put(0,\yy){\line(\xx,0){\xx}}
 \put(\xx,0){\line(0,\yy){\yy}}
 \put(22,144){\makebox(0,0)[t]{\normalsize \bf H\boldmath$_0$ $=$ 59$\pm$2}}
 \put(61,135){\makebox(0,0)[t]{\normalsize \bf H\boldmath$_0$ $=$ 58$\pm$6}}
 \put(10,133){\fbox{\parbox{2.2cm}{\centering SNe\,Ia out to \\
  30\,000\kms}}}
 \put(49.5,124){\fbox{\parbox{2.2cm}{\centering Clusters out to \\
 10\,000\kms}}}
\put(5,83){\fbox{\epsfxsize=3.15cm \epsfbox{tammann.fig1a.eps}}}
\put(45,83){\fbox{\epsfxsize=3.2cm \epsfbox{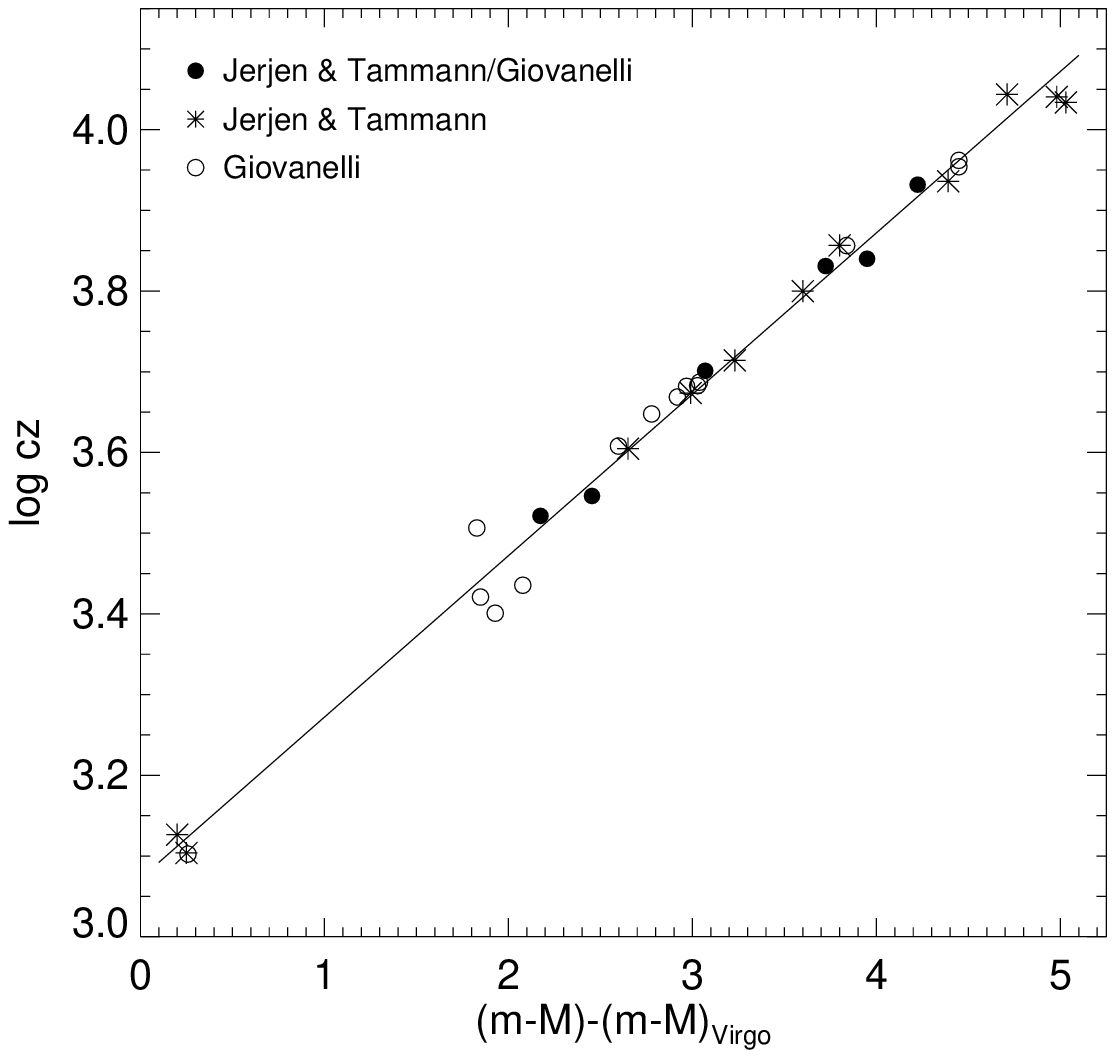}}}
 \put(22,113.5){\vector(0,0){16}}
 \put(62,113.5){\vector(0,0){7}}
 \put(8,78.5){Fig.~1}
 \put(45,58){\fbox{\parbox{3.2cm}{\centering
   Virgo Cluster \\ 
   4 SNe\,Ia \\ 
   $<\!\!{m}^{\rm corr}\!\!>\;=\;$12$\mag$11$\pm$0.15 \\
   $\Rightarrow$ 20.4$\pm$1.5$\;$Mpc}}}
 \put(62,66.5){\vector(0,0){15}}
 \put(83,66){\fbox{\parbox{3.2cm}{\centering
   Fornax Cluster \\ 
   3 SNe\,Ia \\ 
   $<\!\!{m}^{\rm corr}\!\!>\;=\;$12$\mag$24$\pm$0.15 \\
   $\Rightarrow$ 21.7$\pm$1.5$\;$Mpc}}}
 \put(100.5,81){\makebox(0,0)[t]{\normalsize \bf (H\boldmath$_0$ $=$
   62$\pm$10)}}
 \put(31,30){\fbox{\parbox{6cm}{\hspace*{1cm}
   \\ \centering 7 SNe\,Ia \\ 
   $<\!\!{M}^{\rm corr}\!\!>\;=\;-$19$\mag$44$\pm$0.04 \\ \vspace*{0.12cm}}}}
 \put(35,36){\vector(0,0){45.5}}
 \put(62,36){\vector(0,0){15}}
 \put(89,36){\vector(0,0){23}}
 \put(35,10){\fbox{\parbox{2cm}{\centerline{Cepheids}}}}
 \put(67,10){\fbox{\parbox{2cm}{\centerline{Theory}}}}
 \put(46,19){{\sl HST}}
 \put(48,13.5){\vector(1,1){12}}
 \put(77,14.5){\circle*{0.5}}
 \put(76,15.5){\circle*{0.5}}
 \put(75,16.5){\circle*{0.5}}
 \put(74,17.5){\circle*{0.5}}
 \put(73,18.5){\circle*{0.5}}
 \put(72,19.5){\circle*{0.5}}
 \put(71,20.5){\circle*{0.5}}
 \put(70,21.5){\circle*{0.5}}
 \put(69,22.5){\circle*{0.5}}
 \put(68,23.5){\circle*{0.5}}
 \put(68,23.5){\vector(-1,1){2}}
\end{picture}
}
 \end{center}
 \caption{The distance scale built only on SNe\,Ia.}
 \label{fig:2}
\end{figure}

   The 17 high-quality SNe\,Ia {\em inside\/} $10\,000\kms$ give a
value of $H_0$ which is $2-3$ units larger. This is a reflection of
the suspected low-density bubble of corresponding size. 

   An alternate route to $H_0$ is via the Virgo cluster (Fig.~2). Four
well observed SNe\,Ia (excluding the unusual object 1991T) are known
in the Virgo cluster. They give $m_{\rm B}^{\rm corr}=12.11\pm0.15$
and with equation~(14) $(m-M)_{\rm Virgo}=31.55\pm0.16$. 
A miniature Hubble diagram plotting recession velocities versus {\em
  relative\/} cluster distances out to $11\,000\kms$ is shown in
Fig.~2 (cf. e.g. Tamman 1997; Fig.~4 gives a full-size Hubble diagram of the
relative cluster distances). These relative distances are much more 
secure than absolute distances and define the Hubble line with very
small scatter, i.e.
\begin{equation}
  \label{eq:16}
  \log v = 0.2\,[(m-M)_{\rm Cluster} - (m-M)_{\rm Virgo}]+(3.070\pm0.024).
\end{equation}
From equation~(16) follows
\begin{equation}
  \label{eq:17}
  H_0 = -0.2\,(m-M)_{\rm Virgo}+(8.070\pm0.024).
\end{equation}
If the above Virgo cluster modulus from SNe\,Ia is inserted, one finds
$H_0=58\pm6$ at a distance of $\sim\!10\,000\kms$, fully consistent
with equation~(15).

   The three SNe\,Ia in the Fornax cluster provide also a good
distance to this cluster (cf. Fig.~2). However, it is not useful
for the determination of $H_0$ because the observed mean cluster
velocity may be affected by unknown peculiar motions of up to
$\pm20\%$. 

   Some authors have derived higher values of $H_0$ from
SNe\,Ia. These values can be constructed by ($i$) excluding arbitrarily
the brighter calibrators in Table~1, ($ii$) interpreting falsely the
intrinsic color variations of SNe\,Ia as an effect of absorption, and
($iii$) adding ``calibrators'' for which no direct and reliable distance
determinations are available. 

   The conclusion is that SNe\,Ia require $H_0=59$ in perfect
agreement with all independent evidence (cf. Theureau \& Tammann
1998). The internal error is almost vanishingly small. The systematic
error depends on the adopted LMC modulus ($<0\mag1$), the zeropoint
error of the \HST photometry ($<0\mag1$), uncertainties of the
absorption corrections of the calibrators and distant SNe\,Ia
($<0\mag1$), and future improvements of the second-parameter
corrections ($<0\mag1$). It is highly unlikely that these effects
together will affect $H_0$ by more than 10\%. The 95\% confidence
range is therefore $54 \le H_0 < 64$. 

\acknowledgements
The authors thank the Swiss National Science Foundation for financial
support.


\end{document}